\documentclass[aps,prc,twocolumn,groupedaddress,showpacs,showkeys]{revtex4}
\usepackage{graphicx}
\begin{document}
\title{Chemical Equilibrium in Collisions of Small Systems}
\author{I.~Kraus${}^{1,2}$, J.~Cleymans${}^{1,3,4}$, H. Oeschler${}^{1}$, K.~Redlich${}^{4,5}$ and S.~Wheaton${}^{1,3}$}
\affiliation{${}^{1}$Institut f\"ur Kernphysik, Darmstadt University of Technology, D-64289 Darmstadt, Germany\\
	${}^{2}$Nikhef, Kruislaan 409, 1098 SJ Amsterdam, The Netherlands\\
	${}^{3}$UCT-CERN Research Centre and Department  of  Physics,\\ University of Cape Town, Rondebosch 7701, South Africa\\
	${}^{4}$Universit\"at Bielefeld, Fakult\"at f\"ur Physik,  D-33615 Bielefeld, Germany\\
	${}^{5}$Institute of Theoretical Physics, University of Wroc\l aw, Pl-45204 Wroc\l aw, Poland}
\date{\today}
\begin{abstract}
The system-size dependence of particle production in heavy-ion collisions at the top SPS energy is
analyzed in terms of the statistical model.   A systematic comparison is made of two suppression 
mechanisms that quantify strange particle yields in ultra-relativistic heavy-ion collisions:
the canonical model with strangeness correlation radius  determined from the data and the model 
formulated in the canonical ensemble using  chemical off-equilibrium strangeness suppression factor. 
The system-size dependence of the correlation radius and the thermal parameters are obtained  for 
p-p, C-C, Si-Si and Pb-Pb collisions at $\sqrt {s_{NN}} =17.3$ $A$GeV. It is shown that on the basis 
of a consistent set of data there is no clear difference between the two suppression patterns. In the 
present study the strangeness correlation radius was found to  exhibit a rather weak dependence on 
the system size.
\end{abstract}
\maketitle
%
%
\section{Introduction}
Experiments with  ultra-relativistic nucleus-nucleus collisions provide a unique opportunity  to study the
properties of strongly interacting matter under the extreme condition of high energy densities. Hadronic
multiplicities, and their spectra in particular, carry information about the nature of the medium from which they
originated. The statistical model (SM) has been recognized as a powerful approach to describe particle
production yields in heavy-ion collisions ~\cite{hwa,anton}. These models assume that particle creation occurs
at chemical freeze-out with the collision fireball being in thermal and chemical equilibrium.

In the  limit of high temperature and/or  large system size, the grand-canonical (GC) treatment of strangeness
conservation   is adequate in the SM.  There, strangeness conservation is implemented  on  the average and is
controlled by the chemical potential \cite{hwa,anton}. However, if the number of strange particles in a
collision fireball is small, either due to  low temperature (e.g.  in A-A collisions at incident energies of a
few $A$GeV~\cite{Cleymans_99}) or due to  small system size (e.g. in a fireball created in  p-p or  light-ion
collisions), then strangeness conservation has to be implemented exactly  in the canonical (C) ensemble
\cite{hwa,hagedorn,bec1,bec_heinz,hamieh,pbm,marek}. Exact strangeness conservation  leads to the 
suppression of  the strange particle's phase-space and is usually referred to as canonical suppression
\cite{hagedorn,bec_heinz,hamieh}. However, canonical suppression with the assumption of strange\-ness 
chemical equilibrium in the whole fireball volume was found to be insufficient to reproduce the observed yields 
in nucleus-nucleus collisions at the SPS~\cite{marek,Cleymans_99,ckw,steinberg,raf}. In this paper we 
consider a method to account for the suppression beyond the one expected in the canonical ensemble 
by assuming that  exact strangeness conservation holds only in a small sub-volume $V_C$ of the system
\cite{hwa,hagedorn,hamieh}. The concept of such a sub-volume, or a strangeness correlation volume, has 
been used in earlier studies~\cite{redlich,caines,hoehne,bec1}. Here, we present a systematic analysis
of p-p and central C-C, Si-Si and Pb-Pb collisions at the top SPS energy from the NA49
collaboration~\cite{dataCC,dataRos,dataPhi,dataLam,dataBar,dataSi,dataKpi,dataSik,dataPer}.
We selected a consistent  set of data (discussed below) for  different collision systems
and focus only on central nucleus-nucleus collisions. We test the validity of the above
approach and study the relation between $V_C$ and the system size, and also the chemical
freeze-out volume. The results of the canonical model with cluster formation will be
compared to those  obtained in the canonical model with strangeness suppression through
chemical off-equilibrium  factor $\gamma_S$.

The paper is organized as follows: A short outline of the statistical model and its canonical formulation with
cluster formation is given in Section~\ref{SecModel}. In Section~\ref{SecData}, the data analysis is presented
followed by a comparison of the models with the data in Section~\ref{SecRes}. The paper concludes in
Section~\ref{SecDis} with a discussion.
%
%
\section{\label{SecModel}The  model description}
\subsection{Grand-canonical ensemble}
In the statistical model the thermal fireball is characterized by its volume $V$, temperature $T$ and charge  chemical 
potentials $\vec \mu$ which are assumed to be uniform over the whole volume. In the  GC ensemble, these
parameters determine the partition function $Z(T,V,\vec\mu)$. In the hadronic fireball of non-interacting particles and 
resonances, $\ln Z$ is the sum of the contributions of all $i$-particle species of energy $E_i$ and spin-isospin 
degeneracy $g_i$ by,
\begin{equation}
\frac{1}{V} \, {\rm ln}Z(T, V, \vec{\mu}) = \sum_i {}Z_i^1(T, {\vec\mu}), \label{equ1}
\end{equation}
with,
\begin{equation}
Z_i^1= \pm \frac{g_i}{2 \pi^2} \int_0^{\infty} p^2 {\rm d}p ~ {\rm ln}\Bigl[ 1 \pm {\rm
exp} \Bigl( { \frac{{\vec {q_i}} {\vec\mu} -E_i}{T} } \Bigr) \Bigr], \label{equ2}
\end{equation}
where $\vec {q_i}=(B_i,S_i,Q_i)$ are the quantum numbers of particle $i$ and $\vec{\mu}=(\mu_B,\mu_S,\mu_Q)$ are
the chemical potentials related to the conservation of baryon number, strangeness and electric charge, respectively. 
The upper (lower) signs refer to fermions (bosons).

\begin{figure}
\includegraphics[width=0.9\linewidth]{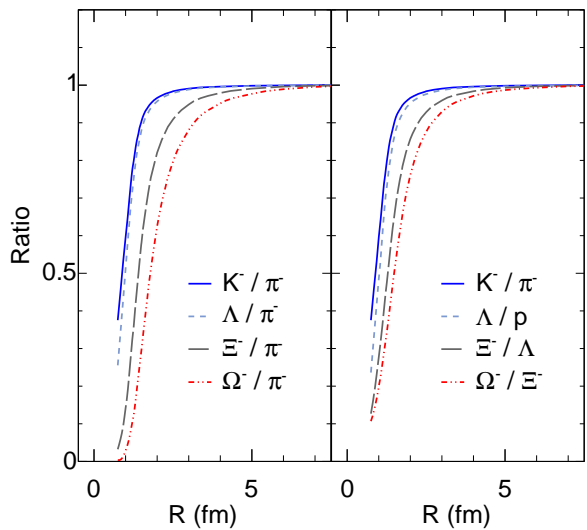}
\caption{Different particle ratios as a function of the radius $R$ of a spherical volume.
The temperature $T$ = 170 MeV and the baryon chemical potential $\mu_B$ = 255 MeV were
chosen according to the thermal conditions at top SPS energy. All ratios are normalized
to their  grand-canonical values. }\label{fig1}
\end{figure}

In a fireball created in heavy-ion collisions  only the baryon chemical potential $\mu_B$ is an independent
parameter while the other two, the charge $\mu_Q$ and the strangeness $\mu_S$  chemical potentials,  are
constrained by the initial conditions from the electric charge of the incoming nuclei and the condition of
strangeness neutrality.

The partition function (\ref{equ1}) contains all information to obtain the number density $n_i$ of thermal
particle species $i$. Introducing the particle's specific chemical potential $\mu_i$, one gets,
\begin{equation}
n_i^{}(T, \vec{\mu}) = \frac{1}{V} \frac{\partial(T {\rm ln}Z)}{\partial
\mu_i}|_{\mu_i=0}. \label{equ3}
\end{equation}
The thermally produced resonances that decay into species $i$ contribute to the measured yield. Therefore, 
the contributions from all heavier particles $j$ that decay to hadron $i$ with the branching fraction 
$\Gamma_{j \rightarrow i}$ have to be calculated as,
\begin{equation}
n_i^{decay} = \sum_j  \Gamma_{j \rightarrow i} ~ n_j. \label{equ4}
\end{equation}
Consequently, the final yield $N_i$ of particle species $i$ is the sum of the thermally
produced particles and the decay products of resonances,
\begin{equation}
N_i = (n_i^{} + n_i^{decay}) ~ V. \label{equ5}
\end{equation}
From Eqs.~(\ref{equ3}-\ref{equ5}) it is clear that in  the GC ensemble the particle yields are determined by the
volume of the fireball, its temperature and the baryon chemical potential.

\subsection{Canonical ensemble}
In general, if the number of particles carrying quantum numbers related to a conservation law is small, then
the grand-canonical description no longer holds. In such a case conservation of charges has to be implemented
exactly in the canonical ensemble. Here, we refer only  to strangeness conservation and
consider charge and baryon number conservation to be fulfilled on the average in the GC
ensemble because the number of charged particles and baryons is much larger than that of strange
particles.
The density of strange particle $i$ carrying strangeness $s$ can be obtained in the canonical ensemble from,

\begin{eqnarray}
n_{i}^C&=&{{Z^1_{i}}\over {Z_{S=0}^C}} \sum_{k=-\infty}^{\infty}\sum_{p=-\infty}^{\infty} a_{3}^{p} a_{2}^{k}
 a_{1}^{{-2k-3p- s}}\nonumber \\
&&I_k(x_2) I_p(x_3) I_{-2k-3p- s}(x_1),   \label{equ6}
\end{eqnarray}
where $Z^C_{S=0}$ is the canonical partition function
\begin{eqnarray}
Z^C_{S=0}&=&e^{S_0} \sum_{k=-\infty}^{\infty}\sum_{p=-\infty}^{\infty} a_{3}^{p}
a_{2}^{k} a_{1}^{{-2k-3p}}\nonumber \\
&& I_k(x_2) I_p(x_3) I_{-2k-3p}(x_1),
\label{eq7a}
\end{eqnarray}
and $Z^1_i$ is the one-particle partition function  (\ref{equ2}) calculated for $\mu_S=0$ in the Boltzmann
approximation. The arguments of the Bessel functions $I_s(x)$ and the parameters $a_i$ are introduced as,
\begin{eqnarray} a_s= \sqrt{{S_s}/{S_{-s}}}~~,~~ x_s = 2V\sqrt{S_sS_{-s}} \label{eq8a}, \end{eqnarray}
where $S_s$ is the sum  of all $Z^1_k(\mu_S=0)$  from (\ref{equ2}) for particle species $k$ carrying strangeness
$s$. 

In the limit where $x_n<1$ (for $n=1$, 2 and 3)   the density of strange particles (\ref{equ6}) carrying
strangeness $s$ is well approximated by~\cite{hwa},
\begin{equation}
n_i^{C} \simeq n_i \frac{I_{s}(x_1)}{I_0(x_1)}. \label{equ7}
\end{equation}
From this  equation it is clear that in the C ensemble the strange particle density depends explicitly on the 
volume $V$ through the arguments of the Bessel functions. In addition, as seen in Eq.~(\ref{equ7}), the
strange particle  densities are   suppressed in the C ensemble due to exact  strangeness conservation 
constraints.

The canonical suppression relative to the GC results  for different particle ratios is illustrated in Fig.~\ref{fig1} 
for fixed $T$ and $\mu_B$  and for different system sizes parameterized by the radius $R$, assuming spherical 
geometry of the volume $V$. The left-hand panel illustrates that the canonical suppression increases with
the strangeness content of the particle \cite{hwa}. For $R > 5$ fm the canonical suppression is already so
small that  the C and GC descriptions of particle ratios are equivalent. The right-hand panel shows the particle
ratios with a difference in their strangeness quantum numbers, $\Delta S$ = 1. Interestingly, the canonical
suppression of these ratios is similar but not identical. Thus, the C-suppression in these ratios  depends on
the strangeness content of both hadrons and not only on the difference.
%
%
\subsection{\label{SecClu}Strangeness suppression mechanisms}

\begin{figure}
\includegraphics[width=0.9\linewidth]{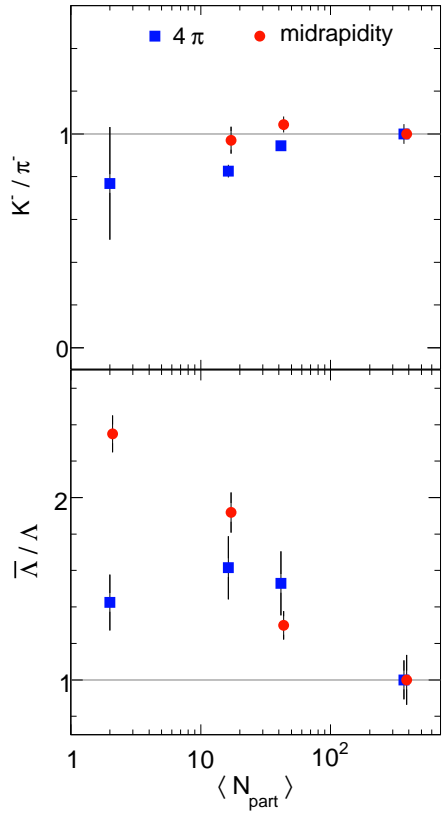}
\caption{ Midrapidity (circles) and 4$\pi$ (squares) data on  the $\bar{\Lambda}/\Lambda$
 and K$^-/\pi^-$ ratios  in p-p and central C-C, Si-Si and Pb-Pb
collisions  normalized to the Pb-Pb measurement \cite{dataCC,dataRos,dataLam,dataBar,dataSi,dataKpi}.}
\label{fig2}
\end{figure}

In the application of the statistical model  to particle production in heavy-ion  and particularly in elementary
particle collisions it was found that canonical suppression alone is not sufficient to quantify the observed
strange particle yields \cite{bec_heinz}. Consequently, additional suppression mechanisms were proposed to
account for deviations from experimental data.
Here we present two methods that lead to additional suppression of strange particle 
phase--space going beyond the normal canonical  effect \cite{hwa,hamieh,bec1}.

The suppression of strangeness has been parameterized by a factor, $\gamma_S$, that is introduced to suppress 
hadrons composed of strange and/or anti-strange quarks \cite{raf}.  In this description the strange particle density
(\ref{equ3}) composed of $s$ strange quarks/antiquarks is modified in the GC ensemble by,
\begin{equation}
n_i\to \gamma_S^s n_i, \label{equ8}
\end{equation}
where $\gamma_S$ is an additional parameter of the model.
This chemical off-equilibrium factor also modifies the calculation of the canonical suppression outlined 
in Eqs. (\ref{equ6}-\ref{eq8a}), the single-particle partition function $Z_k^1$ is replaced by $\gamma_S^s Z_k^1$.

In order to account for an additional strangeness suppression in terms of the canonical model
we use the concept of strangeness correlation in clusters of sub-volume $V_C\leq
V$~\cite{hwa}. Consequently, there are two volume parameters in the model; the overall
volume of the system $V$, which determines the particle yields at fixed density and the
strangeness correlation (cluster) volume $V_C$, which enters through the canonical
suppression factor and reduces the densities of strange particles. 
Hence, this results in replacing the volume $V$ by $V_C$ in Eq. (\ref{eq8a}).
Assuming spherical geometry, the volume $V_C$ is parameterized by the radius $R_C$ which serves as a free
parameter and defines the range of local strangeness equilibrium. A particle with
strangeness quantum number $s$ can appear anywhere in the volume $V$. However, it has to
be accompanied within the sub-volume $V_C$ by other particles carrying strangeness $-s$
to conserve strangeness exactly.

In the application of the SM to particle production and their system-size dependence we consider  three
different formulations of the model:

\begin{description}
\item[(a)] equilibrium model in the canonical  ensemble with strangeness correlation
volume $V_C = V$,

\item[(b)]  non-equilibrium canonical model  with   strangeness undersaturation
parameterized by the factor $\gamma_S$,

\item[(c)]  canonical model that accounts for strangeness correlation in the sub-volume
$V_C \leq V$ that is quantified by the cluster radius $R_C$.
\end{description}
In the following we  compare the above models with experimental data at the top SPS energy for different
colliding systems.  We focus on the system-size dependence of the thermal parameters with particular emphasis on
the change in the strangeness correlation radius $R_C$.

\section{\label{SecData}Data sets}

To verify the statistical nature of particle production in heavy-ion collisions  for different colliding systems
we consider the experimental data at the top SPS energy with $\sqrt{s_{NN}}$~=~17.3~$A$GeV,
from p-p and central C-C, Si-Si and Pb-Pb collisions
\cite{dataCC,dataRos,dataPhi,dataLam,dataBar,dataSi,dataKpi,dataSik,dataPer}. 
In view of the fact that only a limited number of hadrons were experimentally analyzed in the smaller systems, 
we restrict  our study to a consistent set of data~\cite{dataCC,dataRos,dataLam,dataSi,dataKpi}. 
Therefore,  e.g. the multi-strange particles are not included as they are not available for all data sets.
Both midrapidity densities and integrated yields are studied with the exception of the p-p collisions where the
midrapidity data are not available. A compilation of p-p and heavy-ion data used in our analysis is summarized in 
Tables~\ref{Table_pp} and \ref{Table_HI}.

\begin{table}
\caption{\label{Table_pp} Particle yields (4$\pi$ integrated) in minimum bias p-p collisions at
$\sqrt{s_{NN}}$~=~17.3~GeV from Ref.~\cite{dataRos, dataLam} and fit results from model {\bf (c)}.}
\begin{ruledtabular}
\begin{tabular}{l|cc}
&\multicolumn{2}{c}{p-p}\\
Particle &Data&Fit\\
\hline
$\pi^{+}/\pi^{-}$&1.23$\pm$0.26&1.23\\
$K^{+}/K^{-}$&1.61$\pm$0.61&1.65\\
$\bar{\Lambda}/\Lambda$&0.118$\pm$0.013&0.118\\
$K^{-}/\pi^{-}$&0.062$\pm$0.021&0.062\\
$\Lambda/K^{+}$&0.450$\pm$0.138&0.441\\
$\bar{\Lambda}/K^{-}$&0.085$\pm$0.027&0.086\\
$\pi^{-}$&2.57$\pm$0.36&2.57\\
\end{tabular}
\end{ruledtabular}
\end{table}

\begin{table*}
\caption{\label{Table_HI} 4$\pi$-integrated particle yields (upper part) and midrapidity densities (lower part)
in central C-C, Si-Si and Pb-Pb collisions at $\sqrt{s_{NN}}$~=~17.3~GeV and fit results from model {\bf (c)}.
The experimental results were taken from Ref.~\cite{dataCC,dataSi} for C-C and Si-Si and from
Ref.~\cite{dataLam,dataKpi} for Pb-Pb.}
\begin{ruledtabular}
\begin{tabular}{l|cc|cc|cc}
&\multicolumn{2}{c}{C-C}&\multicolumn{2}{c}{Si-Si}&\multicolumn{2}{c}{Pb-Pb}\\
Particle &Data&Fit&Data&Fit&Data&Fit\\
\hline
$\pi^{+}/\pi^{-}$& 1.01 $\pm$ 0.12 & 1.01 & 0.98 $\pm$ 0.11 & 0.98 & 0.97 $\pm$ 0.10 & 0.99\\
$K^{+}/K^{-}$& 1.70 $\pm$ 0.30 & 1.71 & 1.68 $\pm$ 0.26 & 1.70 & 1.98 $\pm$ 0.27 & 1.77\\
$\bar{\Lambda}/\Lambda$& 0.134 $\pm$ 0.039 & 0.134 & 0.127 $\pm$ 0.033 & 0.127 & 0.083 $\pm$ 0.023 & 0.080\\
$K^{-}/\pi^{-}$& 0.067 $\pm$ 0.011 & 0.067 & 0.077 $\pm$ 0.010 & 0.077 & 0.081 $\pm$ 0.010 & 0.082\\
$\Lambda/K^{+}$& 0.52 $\pm$ 0.18 & 0.52 & 0.52 $\pm$ 0.13 & 0.51 & 0.44 $\pm$ 0.10 & 0.51\\
$\bar{\Lambda}/K^{-}$ & 0.119 $\pm$ 0.031 & 0.119 & 0.111 $\pm$ 0.036 & 0.111  & 0.072 $\pm$ 0.015 & 0.072\\
$\pi^{-}$& 22.2 $\pm$ ~1.9 & 22.2 & 57.6 $\pm$ ~4.6 & 57.6 & 639 $\pm$ ~48 & 639\\
\hline
$\pi^{+}/\pi^{-}$& 0.98 $\pm$ 0.21 & 1.07 & 0.99 $\pm$ 0.19 & 1.03 & 0.97 $\pm$ 0.08 & 0.99\\
$K^{+}/K^{-}$& 1.60 $\pm$ 0.28 & 1.30 & 1.60 $\pm$ 0.26 & 1.46 & 1.76 $\pm$ 0.15 & 1.62\\
$\bar{\Lambda}/\Lambda$& 0.269 $\pm$ 0.085 & 0.276 & 0.182 $\pm$ 0.070 & 0.179 & 0.140 $\pm$ 0.044 & 0.122\\
$K^{-}/\pi^{-}$& 0.093 $\pm$ 0.018 & 0.094 & 0.100 $\pm$ 0.018 & 0.101 & 0.096 $\pm$ 0.008 & 0.097\\
$\Lambda/K^{+}$& 0.280 $\pm$ 0.063 & 0.312 & 0.366 $\pm$ 0.081 & 0.396 & 0.361 $\pm$ 0.082 & 0.459\\
$\bar{\Lambda}/K^{-}$& 0.121 $\pm$ 0.034 & 0.112 & 0.107 $\pm$ 0.037 & 0.104 & 0.089 $\pm$ 0.022 & 0.091\\
$\pi^{-}$& 5.70 $\pm$ 0.85 & 5.70 & 15.0 $\pm$ ~2.0 & 15.0 & 175.4 $\pm$ ~~9.7 & 175.4\\
\end{tabular}
\end{ruledtabular}
\end{table*}

The  system-size dependence of ratios of particle yields  is illustrated in Fig.~\ref{fig2}  for the K$^-/\pi^-$
and $\bar{\Lambda}/\Lambda$. The K$^-/\pi^-$ ratio shows rather moderate dependence on the system size.
The midrapidity data on the K$^-/\pi^-$ ratio are almost saturated at the value expected for central Pb-Pb collisions 
already for $N_{\rm part}>20$, whereas the rapidity-integrated ratio shows such a property  only for  $N_{\rm part}>30$.
The variation of the   $\bar{\Lambda}/\Lambda$ ratio with the system size is stronger than that observed in the 
K$^-/\pi^-$ ratio.   At mid-rapidity the $\bar{\Lambda}/\Lambda$  ratio exceeds its value in the full phase-space by 
almost a factor of two.

The yield of $\phi$-mesons per pion in different colliding systems,  normalized to its value in the most central
Pb-Pb collisions, is shown in Fig.~\ref{fig3}. The yield of $\phi$-mesons is seen to decrease stronger with
decreasing system size than  the pion yields. This indicates that the $\phi$  with hidden strangeness does not behave like
a strangeness-neutral object. For comparison, also shown in Fig.~\ref{fig3}, are the results for the system-size
dependence of the normalized $K/\pi$ and   (K/$\pi$)$^2$ ratios. From the comparison with kaons its is not clear
whether the $\phi/\pi$ ratio favors the single- or  double-strange particle properties. Instead, it seems to
behave as a   strange  particle with  an effective quantum number  between one  and two.
%
%
\section{\label{SecRes}Model comparison with  data}

In the SM the particle production yields are fully determined by the thermal conditions in the collision
fireball. In all models {\bf (a)},~ {\bf (b)} and {\bf (c)} there is a common set of parameters $(T,\mu_B,V)$
that quantifies  the thermal particle yields. In the SM models {\bf (b)} and {\bf (c)}, that account for an 
additional strangeness suppression, there are additional free parameters: $\gamma_S$ and $R_C$, 
respectively. For our study, we used the code THERMUS~\cite{thermus}.

\begin{figure}
\includegraphics[width=.9\linewidth]{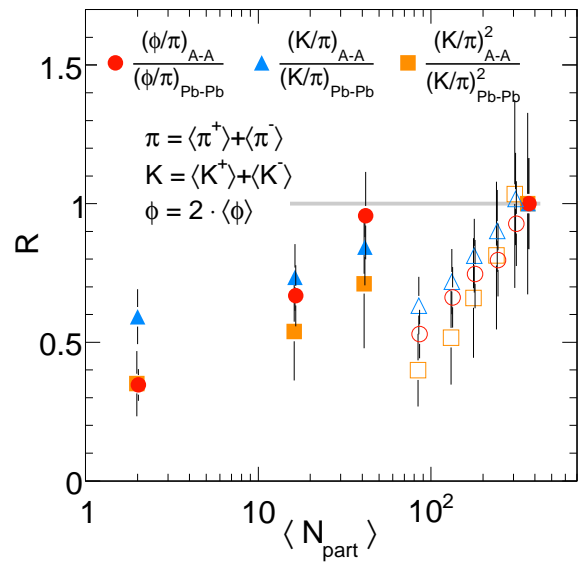}
\caption{ The K/$\pi$, (K/$\pi$)$^2$ and $\phi/\pi$ ratios  of 4$\pi$ integrated yields normalized to their
values in the  central Pb-Pb collisions at the SPS. The p-p and central C-C,  Si-Si   and Pb-Pb data  are from
NA49 Collaboration  at $\sqrt{s_{NN}}$~=~17.3~$A$GeV \cite{dataCC,dataRos,dataPhi,dataSi,dataKpi,dataSik,dataPer}. 
The open symbols refer to non-central Pb-Pb
collisions.}\label{fig3}
\end{figure}

Before considering a detailed quantitative analysis of particle production  for different systems within the SM we
first discuss  general trends expected within the models based on the data shown in Figs.~\ref{fig2} and
\ref{fig3}.

\subsection{General trends}

In terms of the  models   {\bf (a)}, {\bf (b)} and {\bf (c)}  the  anti-strange to strange particle ratios are
independent of  the fireball volume. In the C ensemble an explicit dependence of $n_s$  on the
volume appears in the same way for particles and anti-particles, thus it is cancelled in their ratios.
Consequently, the strong variation of the $\bar\Lambda /\Lambda$ ratio with the system size seen in
Fig.~\ref{fig2} could be  related to the variation in  $T$ and/or $\mu_B$ with the system size. It is quite
natural to assume that the temperature reached in the collision scales approximately with the collision energy.
Therefore, at fixed $\sqrt {s_{NN}}$ the temperature is expected to show a weak  variation with the system size.
On the other hand, the chemical potential can be viewed as the measure of  baryonic stopping in the collision.
Thus it should decrease with $\sqrt {s_{NN}}$ and increase with system size. With the above generic properties of
thermal parameters, the strong variation of $\bar\Lambda /\Lambda$ with the system size seen in Fig.~\ref{fig2} 
could be attributed to a  change in $\mu_B$ at $T\simeq const.$, since in the SM $\bar\Lambda /\Lambda\sim
\exp{(-2\mu_B/T)}$.

In the SM the K$^-$  yield is not explicitly dependent on $\mu_B$. However, due to strangeness neutrality,
$\mu_S=\mu_S(\mu_B)$, implying a weak influence of $\mu_B$ on the kaon yield. Consequently, the change
in the K$/\pi$ ratio with the system size is mainly to be expected as a consequence of canonical suppression
and/or chemical off-equilibrium effects.

\begin{figure}
\includegraphics[width=.8\linewidth]{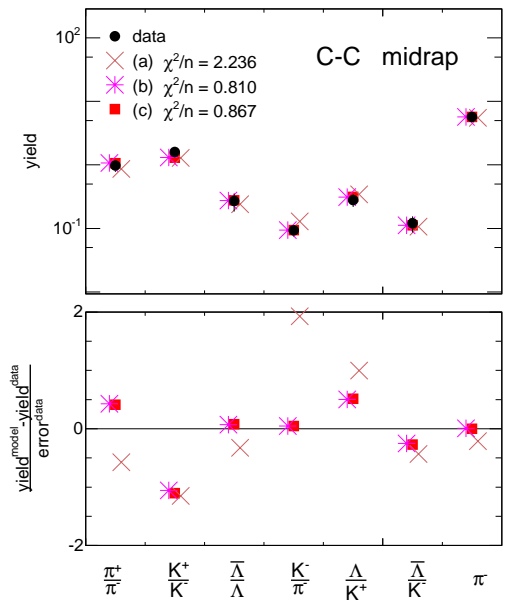}
\caption{Midrapidity $\pi^-$--density and midrapidity particle ratios from C-C collisions at
$\sqrt{s_{NN}}$~=~17.3~$A$GeV from NA49 \cite{dataCC,dataSi}
together with different model predictions: {\bf (a)}, {\bf (b)} and {\bf
(c)} introduced in the text. The lower panel shows the $\chi^2$-deviations of the model fits to
data.}\label{fig4}
\end{figure}
\begin{figure}
\includegraphics[width=.8\linewidth]{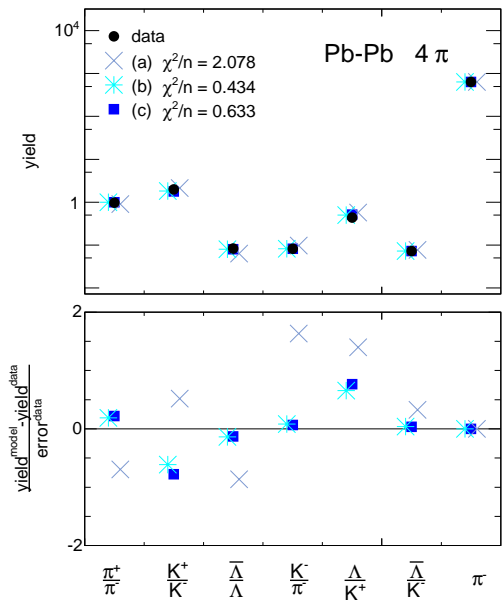}
\caption{ Integrated $\pi^-$ yield and integrated particle ratios from Pb-Pb collisions at
$\sqrt{s_{NN}}$~=~17.3~$A$GeV \cite{dataLam,dataKpi}
together with the model fits as in Fig.~\ref{fig4}. The lower panel shows the
deviation of the model fits to data.}\label{fig5}
\end{figure}
In  the equilibrium statistical  models   {\bf (a)} and {\bf (c)}   the $\phi$-meson, being a strangeness
neutral particle, should show similar system-size dependence as any other non-strange meson. Thus, the strong
variation of the $\phi/\pi$ ratio with centrality seen in Fig.~\ref{fig3} is inconsistent with the SM {\bf
(a)} and {\bf (c)}. The system-size dependence of $\phi$-mesons is only expected to be qualitatively different from  
the other mesons in the model {\bf (b)} that accounts for strangeness suppression by including chemical off-equilibrium 
parameter $\gamma_S$. In this model the $\phi$-meson, being made  of a strange/antistrange quark pair, is suppressed 
by the factor $\gamma_S^2$, thus behaving as a strangeness-two particle. However,  from the discussion in the last 
section it is clear that, on the basis of data, it is not possible to fix the effective strangeness content of the $\phi$-meson.
Consequently, to avoid ambiguities, we have excluded the hidden-strange particles from the SM analysis.
%
%
\subsection{System-size dependence of thermal parameters}
%

\begin{figure}
\includegraphics[width=.8\linewidth]{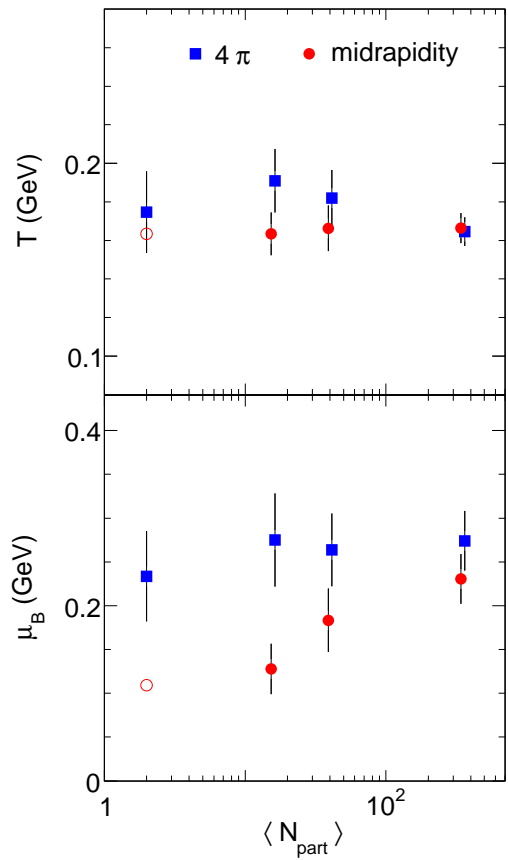}
\caption{Chemical freeze-out temperature $T$  and baryon chemical potential $\mu_B$ from
fits to midrapidity densities (circles) and integrated yields (squares) from p-p and
central C-C, Si-Si and Pb-Pb collisions obtained  in the canonical model  {\bf (c)}
introduced in the text. The open circles refer to the baryon chemical potential at
midrapidity extracted from the $\bar \Lambda/\Lambda$ ratio keeping $T$ as for C-C
collisions.} \label{fig6}
\end{figure}

The bulk properties of the  system-size dependence of different  particle yields in A-A collisions at the top
SPS energy show  qualitative agreement, except for the $\phi$-meson yields, with the expectations of the
statistical models. To verify the   validity of the models requires detailed quantitative studies. In the
following, we analyze the SPS data from p-p, C-C, Si-Si and Pb-Pb collisions in terms of three different
concepts that were  introduced as: {\bf (a)}, {\bf (b)} and {\bf (c)}  in the previous section. The
investigated yields and ratios  (listed in Tables~\ref{Table_pp} and \ref{Table_HI}) together with model
predictions are shown in  Figs.~\ref{fig4} and \ref{fig5} for C-C and Pb-Pb collisions respectively. Particle
ratios from the fits with the best $\chi^2$ are displayed for all of the models. The model {\bf (a)} with a
completely equilibrated strangeness abundance exhibits large deviations from  data.  The normalized differences
between the model results and the data are shown in the lower panels of these figures. For the small systems,
model {\bf (a)} features very large values of $\chi^2\sim 2$ per degree of freedom.
Thus, the equilibrium SM with strangeness  conservation in the C ensemble is not favored by the data.   In
contrast, both models that allow for an extra suppression of the strange-particle phase-space  with $\gamma_S$
or with $R_C$, agree quite well with the measurements and yield comparably good descriptions of data. The
latter can be correlated to properties of the hot and dense medium and systematic studies might provide access
to the production and hadronization mechanism of strange particles.

\begin{figure}
\includegraphics[width=.9\linewidth]{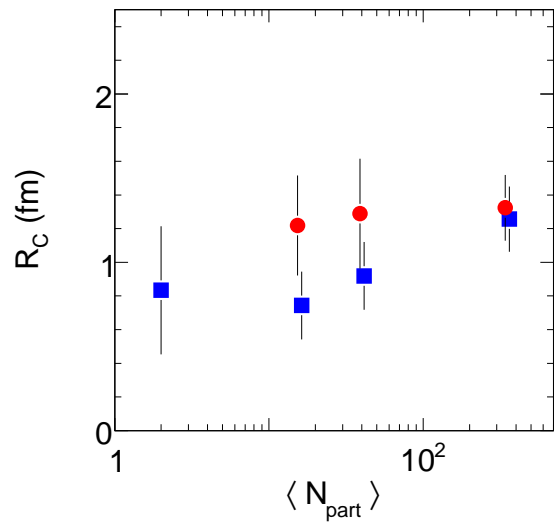}
\caption{Cluster radius $R_C$ as a function of the system size in A-A collisions  from fits of the model {\bf
{\bf (c)}} to midrapidity densities (circles) and integrated yields (squares) from p-p and central C-C, Si-Si
and Pb-Pb collisions.} \label{fig7}
\end{figure}

\begin{figure}
\includegraphics[width=.9\linewidth]{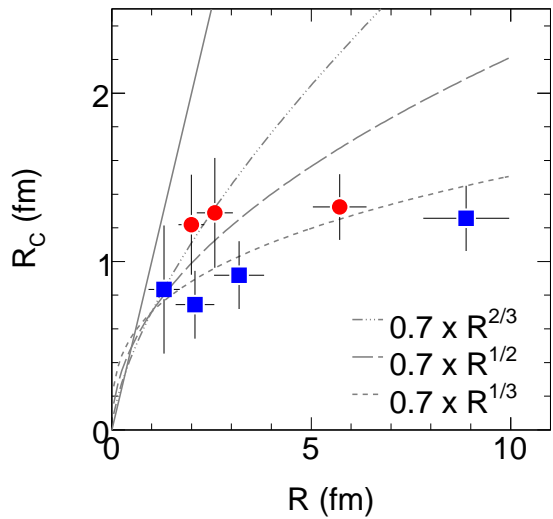}
\caption{Cluster radius $R_C$ as a function of the fireball radius $R$ from fits of the model {\bf (c)} to
midrapidity densities (circles) and integrated yields (squares) from p-p and central C-C, Si-Si and Pb-Pb
collisions. The broken lines show various choices of the $R_C = c \, R^{\alpha}$ relation. The full line
indicates the proportionality $R_C = R$.} \label{fig8}
\end{figure}

The system-size dependence of thermal parameters obtained in these studies within the models {\bf (a)} and {\bf
(b)} agree with previously published results and thus will not be discussed here. Instead, in Figs.~\ref{fig6},
\ref{fig7} and \ref{fig8} we concentrate on the system-size dependence of thermal parameters characterizing  the
canonical model {\bf (c)} with cluster formation. The corresponding values are summarized  in Table~\ref{Table_fit}.

\begin{table*}
\caption{\label{Table_fit} Statistical model parameters, extracted from the comparison of model {\bf (c)} with
experimental data from full phase-space (4$\pi$ integrated) (upper part) and midrapidity densities (lower part)
in minimum bias p-p, central C-C, Si-Si and Pb-Pb collisions at $\sqrt{s_{NN}}$~=~17.3~GeV.}
\begin{ruledtabular}
\begin{tabular}{lcccc}
Reaction & $T \rm{(MeV)}$ &  $\mu_B \rm{(MeV)}$ &  $R \rm{(fm)}$ &  $R_C \rm{(fm)}$ \\
\hline
p-p&174.7$\pm$21.2&234$\pm$51&1.31$\pm$0.39&0.83$\pm$0.38\\
C-C&191.0$\pm$16.5&275$\pm$53&2.09$\pm$0.49&0.74$\pm$0.20\\
Si-Si&181.9$\pm$14.6&264$\pm$42&3.19$\pm$0.63&0.92$\pm$0.20\\
Pb-Pb&164.6$\pm$~7.5&274$\pm$34&8.9$\pm$1.1&1.26$\pm$0.19\\
\hline
C-C&163.4$\pm$11.1&128$\pm$29&1.99$\pm$0.32&1.22$\pm$0.30\\
Si-Si&166.3$\pm$11.8&183$\pm$36&2.58$\pm$0.46&1.29$\pm$0.33\\
Pb-Pb&166.4$\pm$~7.8&231$\pm$28&5.71$\pm$0.67&1.32$\pm$0.20\\
\end{tabular}
\end{ruledtabular}
\end{table*}

The freeze-out temperature $T$ extracted from midrapidity data is barely dependent on the system size, as shown
in Fig.~\ref{fig6}. This is in agreement with the expectation that the chemical freeze-out temperature is predominantly 
established by the collision energy. Within statistical errors the $4\pi$ data are also consistent with having the same
chemical freeze-out temperature for all  colliding systems from p-p up to Pb-Pb.

No significant change in the baryon chemical potential is seen in Fig.~\ref{fig6}  for $4\pi$ data. The
variation of $\bar\Lambda/\Lambda$ seen in the $4\pi$ data in Fig.~\ref{fig2} is a combined effect  of  minor
changes in the freeze-out temperature and the chemical potential in  different systems. At midrapidity, the
chemical potential shows a decrease by almost 100 MeV from central Pb-Pb to C-C collisions at constant
temperature. Consequently, there is a steep increase of the $\bar\Lambda/\Lambda$  ratio, as seen in
Fig.~\ref{fig2}.

From the above results it is clear that at $y=0$ the collision fireball created in A-A
collisions appears at constant temperature but with varying $\mu_B$ that  decreases with
decreasing $N_{\rm part}$. For the rapidity integrated data such systematics cannot be
concluded.

In  the canonical model {\bf (c)} there are two volume scales that characterize the system: the fireball radius
at freeze-out $R$ and  the strangeness correlation radius $R_C$. The fireball radius $R$ is determined by the
particle multiplicity e.g. the pion yield. Thus, a smaller pion density at  midrapidity   causes a smaller
freeze-out radius than for the integrated yields.


The strangeness correlation volume $R_C$ is extracted from the system-size dependence of the strange to 
non-strange particle ratios. The resulting system-size dependence of $R_C$ is shown in Fig.~\ref{fig7}. An 
increase of single-strange to non-strange particle ratios  from p-p to Pb-Pb reactions is reflected in a variation 
of the strangeness correlation volume with the system size. The cluster radius varies between 0.7 fm and 1.3 
fm. A larger radius in the fits of midrapidity data is correlated with the experimental observation of larger 
$K^+/\pi^+$ ratios at midrapidity as compared to $4\pi$ yields.

\section{\label{SecDis}Discussion}

We have already discussed in Fig.~\ref{fig1} that for strangeness-one particles  the
canonical suppression for $R>(2-3)$ fm shows a very flat change with $R$ and is hardly
distinguishable from  the asymptotic GC-value. Consequently, an exact determination of
$R_C$ based only on single-strange   particles is very difficult and requires  high
precision data. For $R>3$ fm an increase of $R_C$ by a large factor will not influence
the  particle yields significantly. Therefore, the results shown in Figs.~\ref{fig7} and
\ref{fig8} for $N_{\rm part}>30$ or $R>5$ fm should be considered as a lower limit. The
actual change of $R_C$ with system size and the chemical freeze-out radius could be
stronger than that shown in Figs. ~\ref{fig7} and ~\ref{fig8}. This happens, for example, at the
SIS energy where the correlation volume $V_C$ was shown to scale with the number of
projectile participants \cite{Cleymans_99}. A more precise determination of the relation
between the volume at chemical freeze-out, system size and strangeness correlation at the
SPS would require the analysis of the system size and/or centrality dependence  of
multi-strange particles in A-A collisions that show a stronger sensitivity to the
strangeness correlation volume \cite{centrality,caines,stock}.

Relating  the cluster  to the fireball radii $R_C=R_C(R)$,  it becomes clear that R$_C$ has a significantly
weaker system-size dependence than $R$, as seen in  Fig.~\ref{fig8}. In p-p collisions the strangeness
correlation length  $R_C$ is almost as large as the fireball radius. In contrast, with increasing system size,
the cluster volume  can be even a few times smaller than the entire fireball. The dependence of $R_C$ on $R$
shown in Fig.~\ref{fig8} can  be parameterized as $R_C=c\cdot R^\alpha$ with $\alpha <1$ and $c\simeq 0.7$ fm.
From the consistent data set used in these studies the value of $\alpha$ appears  in the range $1/3<\alpha<1/2$.
From this analysis we could conclude that the transition from suppressed (in p-p) to almost saturated (in Pb-Pb)
single-strange particle  production happens in a  narrow range of the correlation length. This  could indicate
that $R_C \simeq$ (1-2) fm is the length scale of strangeness correlation. This suggestion gets implicit support
from the percolation model calculations~\cite{satz}.

The appearance of a particle correlation volume in p-p collisions, introduced here for
strangeness, comes also from other observables. The mass shift of the $\rho^0$-meson in
the $\pi^+ \pi^-$ decay channel observed in p-p collisions~\cite{starrho} could not  be
substantially accounted for by the phase-space distortion of a Breit-Wigner line shape.
It was argued~\cite{fachini} that re-scattering and re-interaction of pions contribute an
additional term to the $\rho^0$ spectral function. The best agreement with the
measurement is achieved if a range of $\pi^+ \pi^-$ scattering of 0.73~fm is chosen.
This value is in  agreement with the strangeness correlation length  $R_C$  obtained in
this study  from the statistical model analysis of p-p  data at the SPS.

In summary, two modified versions of the canonical statistical model have been investigated.
The experimentally observed strong suppression of the strange-particle phase-space is retraced to local
strangeness conservation within small correlated clusters in the fireball. These sub-volumes, in consequence,
cause strong canonical suppression and allow one to successfully reproduce the experimental data. Furthermore, the
cluster size is found to be only weakly dependent on system size. In all data under study, it is of the order of
1 to 2~fm.

 
\section*{ Acknowledgments}
 K.R.  and J.C. acknowledge a   partial support of the DFG under grant GRK 881. K.R. also
 acknowledges
the support of  the Polish Ministry of National Education (MEN).

%
%


\begin{thebibliography}{99}

\bibitem{hwa} P.~Braun-Munzinger, K.~Redlich and J.~Stachel,  {\sl Quark-Gluon Plasma 3},
edited by R. C. Hwa and Xin-Nian Wang (World Scientific Publishing, 2003),
nucl-th/0304013.
%
\bibitem{anton} A.~Andronic, P.~Braun-Munzinger and J.~Stachel,
Nucl.\ Phys.\  A {\bf 772}, 167 (2006)  [arXiv:nucl-th/0511071].
%
\bibitem{Cleymans_99}J. Cleymans, H. Oeschler, K. Redlich, Phys. Rev. C{\bf 59} 1663 (1999).
%
\bibitem{hagedorn} R.~Hagedorn and K.~Redlich,
Z.\ Phys.\  C {\bf 27}, 541 (1985).
%
\bibitem{bec1}       F.~Becattini, Z. Phys. C69 (1995) 485. 
%
\bibitem{bec_heinz}  F.~Becattini and U.~Heinz, Z. Phys. C76 (1997) 269.
%
\bibitem{hamieh}J. S. Hamieh, K. Redlich and A. Tounsi, Phys. Lett. { B486} (2000) 61.
%
\bibitem{pbm} P.~Braun-Munzinger, J.~Cleymans, H.~Oeschler, K.~Redlich, Nucl. Phys. A 697 (2002) 902.
%
\bibitem{marek} F.~Becattini, J.~Manninen and M.~Gazdzicki,
Phys.\ Rev.\  C {\bf 73}, 044905 (2006);
F.~Becattini, J.~Cleymans, A.~Keranen, E.~Suhonen and K.~Redlich,
Phys.\ Rev.\  C {\bf 64}, 024901 (2001).
%
\bibitem{raf}       J.~Rafelksi and A.~Tounsi, Phys. Lett. B \textbf{292} (1992) 417;
J. Letessier, J. Rafelski, A. Tounsi, Phys. Rev. C {\bf 50}, 406 (1994); C. Slotta, J.
Sollfrank, U. Heinz, Proc. of Strangeness in Hadronic Matter (Tucson), (Ed.\ J.
Rafelski), AIP conference proc.\ {\bf 340} (1995) p.\ 462.
%
\bibitem{ckw} J.~Cleymans, B.~K\"ampfer, S.~Wheaton, Phys. Rev. C 65 (2002) 027901.
%
\bibitem{steinberg} J.~Cleymans, B.~K\"ampfer, P.~Steinberg, S.~Wheaton, J. Phys. G 30 (2004) S595.
%
\bibitem{redlich} A.~Tounsi, A.~Mischke, K.~Redlich, Nucl. Phys. A715 (2003) 565.
%
\bibitem{caines} H.~Caines, J. Phys. G {\bf 32} (2006) S171.
%
\bibitem{hoehne} C.~H\"ohne, F.~P\"uhlhofer, R.~Stock, Phys. Lett. B \textbf{640} (2006) 96.
%
\bibitem{dataCC}        I.~Kraus (NA49 Collaboration), J. Phys. G {\bf 31} S147 (2005), [nucl-ex/0412050].
%
\bibitem{dataRos}   A.~M.~Rossi et al., Nucl. Phys. B {\bf 84} 269 (1975) .
%
\bibitem{dataPhi}       S.~V.~Afanasiev et al. (NA49 Collaboration), Phys. Lett. B {\bf 491} 59 (2000).
%
\bibitem{dataLam}   T.~Anticic et al. (NA49 Collaboration), Phys. Rev. Lett. {\bf 93} 022302 (2004), [nucl-ex/0311024].
%
\bibitem{dataBar}   D.~Barna (NA49 Collaboration), Ph.D. Thesis, University and KFKI Budapest, Hungary (2002).
%
\bibitem{dataSi}        C.~Alt et al. (NA49 Collaboration), Phys. Rev. Lett. {\bf 94} 052301 (2005), [nucl-ex/0406031].
%
\bibitem{dataKpi}       S.~V.~Afanasiev et al. (NA49 Collaboration), Phys. Rev. C {\bf 66} 054902 (2002).
%
\bibitem{dataSik}       F.~Sikler (NA49 Collaboration), Nucl. Phys. A {\bf 661} 45c (1999).
%
\bibitem{dataPer}   V.~Friese (NA49 Collaboration), Nucl. Phys. A {\bf 698} 487c (2002).
%
\bibitem{thermus}   S.~Wheaton and J.~Cleymans, J.~Phys. G31 (2005) S1069.
%
\bibitem{stock}     R.~Stock, Phys. Lett. B \textbf{456} (1999) 277.
%
\bibitem{centrality}    A.~Tounsi and K.~Redlich, J. Phys. G {\bf 28} S2095 (2002).
%
\bibitem{satz}      M.Nardi and H.~Satz, Phys. Lett. B {\bf 442} 14 (1998).
%
\bibitem{starrho}       J.~Adams et al. (STAR Collaboration), Phys. Rev. Lett. {\bf 92} 092301 (2004).
%
\bibitem{fachini}       P.~Fachini, R.~S.~Longacre, Z.~Xu and H.~Zhang, J.~Phys. G {\bf 33} (2007) 431.
%
\end{thebibliography}
\end{document}